\documentclass[12pt]{article}

\usepackage{amssymb}
\usepackage{amsmath}
\usepackage{amscd}
\usepackage{latexsym}
\usepackage{graphicx}

\usepackage{cite}

\newcommand{\be}{\begin{equation}}
\newcommand{\ee}{\end{equation}}

\newcommand{\dlt}{\delta}
\newcommand{\prt}{\partial}
\newcommand{\br}{{\bf r}}

\newcommand{\vp}{\varphi}
\newcommand{\ep}{\varepsilon}
\newcommand{\al}{\alpha}
\newcommand{\ra}{\rightarrow}
\newcommand{\sgm}{\sigma}

\newcommand{\gm}{\gamma}
\newcommand{\om}{\omega}

\newcommand{\Gm}{\Gamma}
\newcommand{\dgr}{\dagger}

\newcommand{\rgl}{\rangle}
\newcommand{\lgl}{\langle}

\begin{document}

\begin{center}
{\Large {\bf Some typical delusions in the theory of Bose-Einstein condensation
} \\ [5mm]

V.I. Yukalov$^{1,2}$ } \\ [3mm]

{\it $^1$Bogolubov Laboratory of Theoretical Physics, \\
Joint Institute for Nuclear Research, Dubna 141980, Russia \\ [2mm]

$^2$Instituto de Fisica de S\~ao Carlos, Universidade de S\~ao Paulo, \\
CP 369, S\~ao Carlos 13560-970, S\~ao Paulo, Brazil} \\ [3mm]

{\bf E-mail}: yukalov@theor.jinr.ru

\end{center}

\vskip 2cm

\begin{abstract}
Despite the long history of the theory of Bose-Einstein condensation, there exist till 
nowadays some slippery points that are often misunderstood and result in confusion. 
The report touches some of these points, explaining the following: Global gauge 
symmetry breaking is the necessary and sufficient condition for the existence of 
Bose-Einstein condensate. There is no any ``grand canonical catastrophe". The stability 
of the ideal Bose gas depends on the spatial dimensionality and the shape of a trap. 
Symmetry-broken averages cannot be neglected. The so-called ``Popov approximation", 
ascribed to Popov, suggesting to neglect anomalous averages, is neither an approximation 
nor has anything to do with Popov. There are no thermodynamically anomalous fluctuations 
in stable equilibrium systems. Representative statistical ensembles are equivalent. 
\end{abstract}

\vskip 1cm

{\bf Introduction}. Although the theory of Bose-Einstein condensation (BEC) enjoys a 
long history, there remains till nowadays some subtle points that, to a great surprise, 
are poorly understood. The current literature on BEC is full of wrong statements, 
incorrect applications of theory, and erroneous results. In the present report, we touch 
some of the mostly spread delusions in the BEC theory. Explanations will follow the style 
of a report, because of which they will be rather concise. A more detailed treatment is 
planned for a review that, hopefully, will appear soon.

\vskip 2mm

{\bf Criterion of Bose-Einstein condensation}. In the second-quantization representation,
the system is characterized by the field operators $\psi$ that can be expanded in natural 
orbitals,
\be
\label{1}
 \psi(\br) \; = \; \psi_0(\br) + \psi_1(\br)  \; = \;
a_0 \vp_0(\br) + \sum_{k\neq 0} a_k \vp_k(\br) \;  ,
\ee
where the first term describes the field operator of particles in the lowest level 
\cite{Penrose_1}. In a finite system, strictly speaking, there is no BEC, but for its 
correct characterization, one needs to consider the thermodynamic limit that for an
untrapped system has the standard form
\be
\label{2}
N  \; \ra \;  \infty \; , \qquad V  \; \ra \; \infty \; , \qquad
\frac{N}{V}  \; \ra \; const \; .
\ee
For trapped systems, the thermodynamic limit is generalized \cite{Yukalov_2} to the form
\be
\label{3}
N  \; \ra \;  \infty \; , \qquad A_N  \; \ra \; \infty \; , \qquad
\frac{A_N}{N}  \; \ra \; const \;   ,
\ee   
in which $A_N$ is an extensive quantity. The criterion of the condensate existence reads 
as
\be
\label{4}
 \lim_{N\ra\infty} \; \frac{\lgl\; a_0^\dgr \; a_0 \; \rgl}{N} \; > \; 0 \;  ,
\ee
where the thermodynamic limit is assumed. 

\vskip 2mm
{\bf Global gauge symmetry breaking}. As many other phase transitions, the BEC 
transition is accompanied by symmetry breaking. This is the global gauge symmetry 
breaking, when there appear nonzero symmetry-broken anomalous averages. Thus the 
field-operator statistical average
\be
\label{5}
\lgl\; \psi(\br) \; \rgl \; = \; \eta(\br) 
\ee
gives the condensate wave function that is an order parameter of the symmetry-broken 
phase. The other example is
$$
\lgl\; \psi(\br) \; \psi(\br') \; \rgl \; = \; \eta(\br)\; \eta(\br') +
\lgl\; \psi_1(\br) \; \psi_1(\br') \; \rgl \;  .
$$   
To break the system gauge symmetry, it is possible to use the method of infinitesimal 
sources that is also called the method of quasi-averages. Then the initial gauge-invariant 
system Hamiltonian $H$ is supplemented with a source breaking the symmetry, thus obtaining
\be   
\label{6}
H_\ep \; = \; H + \ep \hat B \; .
\ee
The averages of the operators representing extensive observable quantities are defined 
by the rule
\be
\label{7}
\lim_{\ep\ra 0} \; \lim_{N\ra\infty} \;
\frac{\lgl\; \hat A_N \; \rgl_\ep}{N} \;  ,
\ee
where the statistical averaging employs the symmetry-broken Hamiltonian (\ref{6}).   

Bogolubov \cite{Bogolubov_3,Bogolubov_4,Bogolubov_5} and Ginibre \cite{Ginibre_6} 
proved that the spontaneous gauge symmetry breaking implies the appearance of BEC and 
that the symmetry breaking by means of the infinitesimal sources is equivalent to the 
symmetry breaking by the canonical transformation, called the Bogolubov shift,
\be
\label{8}
 \psi(\br) \; = \; \eta(\br) + \psi_1(\br) \;  ,
\ee
where $\eta$ is a condensate wave function and $\psi_1$ is the field operator of 
non-condensed particles. Roepstorff \cite{Roepstorff_7} proved that the occurrence of 
BEC is necessarily accompanied by the gauge symmetry breaking,
\be
\label{9}
\lim_{N\ra\infty} \; \frac{\lgl\; a_0^\dgr \; a_0 \; \rgl}{N} \; \leq \;
\lim_{\ep\ra 0} \; \lim_{N\ra\infty} \;
\frac{|\; \lgl\; a_0 \; \rgl_\ep\; |^2}{N} \; .
\ee
In that way, it is a rigorously proved mathematical fact that: {\it Spontaneous gauge 
symmetry breaking is a necessary and sufficient condition for the Bose-Einstein 
condensate existence}. As soon as the gauge symmetry is broken, it is necessary to take
into account the existence of symmetry-broken anomalous averages. The use of the
approximations not taking account of these averages, such as the Hartree or Hartree-Fock
approximations, is incorrect.  

\vskip 2mm
{\bf Conditions of thermodynamic stability}. Statistical systems can exist in equilibrium  
states provided they are stable. Formally, it may look that BEC can occur in a system, but
actually, being unstable, it cannot exist in equilibrium. One of the most important 
stability conditions for BEC is the restriction on the isothermic compressibility
\be
\label{10}
 0 \; \leq \; \varkappa_T \; < \; \infty \;  ,   
\ee
where the compressibility is  
\be
\label{11}
\varkappa_T \; = \; - \;
\frac{1}{V} \;\left( \frac{\prt V}{\prt P} \right)_{TN}  \; = \;
\frac{1}{\rho N} \;\left( \frac{\prt N}{\prt \mu} \right)_{TV}  \;  .
\ee
Since the compressibility can be represented in the form
\be
\label{12}
\varkappa_T \; = \; \frac{{\rm var}(\hat N)}{\rho T N} \; ,
\ee
where the variance is
$$
{\rm var}(\hat N) \; \equiv \; 
\lgl \; \hat N^2 \; \rgl - \lgl \;\hat N \; \rgl^2 \; ,
$$
the stability condition (\ref{10}) is equivalent to the condition on the particle 
relative variance
\be
\label{13}
0 \; \leq \;  \frac{{\rm var}(\hat N)}{N} \; < \; \infty \; .
\ee
     
\vskip 2mm
{\bf Myth of ``grand canonical catastrophe"}. A very popular nonscientific fiction 
is the occurrence of the so-called ``grand canonical catastrophe". One claims that the
particle fluctuations of condensate, in the frame of the grand canonical ensemble, are
characterized by the variance ${\rm var}({\hat N}_0) \propto N_0(1+N_0)$. Hence, in the 
presence of the condensate, when $N_0 \propto N$, the particle fluctuations are 
catastrophic, being of order $N^2$. Such catastrophic fluctuation, if they would really 
exist, would make the system strongly unstable. One even concludes that the grand 
canonical ensemble is not equivalent to canonical and, in principle, cannot be applied 
to Bose-condensed systems. This conclusion, however, is caused by the principal mistake
where one uses the grand canonical ensemble without gauge symmetry breaking, while the
BEC is present. If the gauge symmetry is not broken, this has to do with the normal
non-condensed system, where $N_0$ cannot be proportional to $N$. But as soon as BEC 
occurs, and $N_0$ becomes proportional to $N$, the gauge symmetry becomes broken and 
the condensate variance is not proportional to $N_0(1+N_0)$. Then no catastrophic 
fluctuations arise.  

Instead, in the presence of the BEC, one has to consider the grand canonical ensemble
with the broken gauge symmetry. This can be done by adding to the system Hamiltonian 
a symmetry-breaking term, say in the form
\be
\label{14}
\ep \; \hat B \; = \; \ep \; \sqrt{V} \; \left( a_0^\dgr e^{i\vp} +
a_0 e^{-i\vp} \right) \;   .
\ee
Then, following the Bogolubov method of quasi-averages, it is straightforward to find
\be
\label{15}
\lim_{\ep\ra 0} \; \lim_{N\ra\infty} \;
\frac{{\rm var}(\hat N_0)}{N} \; = \; 0 \;   .
\ee
This tells us that the condensate fluctuations are neither catastrophic nor exist at al. 
The same conclusion can be obtained in even a simpler way, by employing the Bogolubov
shift (\ref{8}) giving ${\hat N}_0 = |\eta|^2$. This immediately leads to 
${\rm var}(N_0) \equiv 0$. See details in \cite{Yukalov_9}. Thus, the particle 
fluctuations are due solely to the fluctuations of non-condensed particles,      
${\rm var}({\hat N}) = {\rm var}({\hat N}_1)$. 

\vskip 2mm
{\bf Uniform ideal Bose gas}. The stability of a system, characterized by a finite 
compressibility, hence a finite relative particle variance, depends of the spatial 
dimensionality and the shape of a trap \cite{Yukalov_9}. For this purpose, we have 
to study the particle scaling of the relative variance defining the compressibility 
(\ref{12}). 

Let us consider an ideal uniform Bose gas, or the gas in a box-shaped trap. For 
$T > T_c$, the variance, hence compressibility for $d \leq 2$ is negative, implying 
instability,
$$
\frac{{\rm var}(\hat N)}{N} \; \propto \; - \; \frac{1}{N} \qquad 
( d \leq 2) \; ,
$$
\be
\label{16}
 \frac{{\rm var}(\hat N)}{N} \; \propto \; const \qquad 
( d > 2) \qquad ( T > T_c) \;  .
\ee
The gas is stable for the dimensionality $d > 2$.

After BEC has occurred at $T < T_c$, the particle fluctuations scale as
$$
\frac{{\rm var}(\hat N)}{N} \; \propto \; N^3 \qquad 
( d = 1) \; ,
$$
$$
\frac{{\rm var}(\hat N)}{N} \; \propto \; N \qquad 
( d = 2) \; ,
$$
$$
\frac{{\rm var}(\hat N)}{N} \; \propto \; N^{1/3} \qquad 
( d = 3) \; ,
$$
$$
\frac{{\rm var}(\hat N)}{N} \; \propto \; \ln N \qquad 
( d = 4) \; ,
$$
\be
\label{17}
 \frac{{\rm var}(\hat N)}{N} \; \propto \; const \qquad 
( d > 4) \qquad ( T < T_c) \;   .
\ee
Thence, the gas is stable for $d > 4$, but for the lower dimensions it is unstable. 
   
\vskip 2mm
{\bf Trapped ideal Bose gas}. Bose gases are trapped in power-law potentials of the form
\be
\label{18}
 U(\br) \; = \; \sum_{\al=1}^d \frac{\om_\al}{2} \; 
\left| \; \frac{r_\al}{l_\al} \; \right|^{n_\al}\;  .
\ee
The effective frequency and length are defined by the expressions
\be
\label{19}
\om_0 \; \equiv \; \left( \; \prod_{\al=1}^d \om_\al \; \right)^{1/d} \; ,
\qquad
l_0 \; \equiv \; \left( \; \prod_{\al=1}^d l_\al \; \right)^{1/d} \; .
\ee
The important quantity is the effective confining dimension \cite{Yukalov_9}
\be
\label{20}
 D \; \equiv \; \frac{d}{2} + \sum_{\al=1}^d \frac{1}{n_\al} \; .
\ee
For defining the effective thermodynamic limit, we can take the average system energy
\be
\label{21}
 E_N \; = \; \frac{D}{\gm_D} \; g_{D+1}(z) \; T^{D+1} \;  ,
\ee
in which
\be 
\label{22}
\gm_D \; \equiv \; \frac{\pi^{d+2}}{2^D} \; \prod_{\al=1}^d
\frac{\om_\al^{1/2+1/n_\al}}{\Gm(1+1/n_\al)} \;  ,
\ee
thus considering the limit
\be
\label{23}
N \; \ra \; \infty \; , \qquad E_N \; \ra \; \infty \; , \qquad 
\frac{E_N}{N} \; \ra \; const \;  .
\ee
The latter takes the form of the limit
\be
\label{24}
N \; \ra \; \infty \; , \qquad \gm_D \; \ra \; 0 \; , \qquad 
N \gm_D \; \ra \; const \;   .
\ee

Above $T_c$, the relative particle variance scales as
$$
\frac{{\rm var}(\hat N)}{N} \; \propto \; - \; \frac{1}{N}
\qquad ( D \leq 1 ) \; ,
$$
\be
\label{25}
\frac{{\rm var}(\hat N)}{N} \; \propto \; const
\qquad ( D > 1 ) \qquad ( T > T_c) \;   ,
\ee
which implies that the gas is stable only for $D > 1$.  
    
Below $T_c$, the scaling is as follows:
$$
\frac{{\rm var}(\hat N)}{N} \; \propto \; N^{(2-D)/D}
\qquad ( D < 2 ) \; ,
$$
$$
\frac{{\rm var}(\hat N)}{N} \; \propto \; \ln N
\qquad ( D = 2 ) \; ,
$$
\be
\label{26}
\frac{{\rm var}(\hat N)}{N} \; \propto \; const
\qquad ( D > 2 ) \qquad ( T < T_c) \;   .
\ee
Hence the gas is stable for $D > 2$.

\vskip 2mm
{\bf Conserving versus gapless approaches}. Hohenberg and Martin \cite{Hohenberg_10} 
noticed that for a Bose system with broken gauge symmetry there always happens the 
dilemma of conserving versus gapless approaches. This dilemma exhibits itself by either 
leading to a not self-consistent approach where thermodynamic relations are not valid, 
or resulting in a gapful spectrum, while, according to the Hugenholtz--Pines theorem
\cite{Hugenholtz_11}, the spectrum has to be gapless. See a detailed discussion in Ref. 
\cite{Yukalov_12}. 

However, this dilemma does not arise when using a representative ensemble taking into 
account all conditions imposed on the system. Above the critical temperature of BEC, 
there exists one kind of particles and one condition for the total average number of 
particles $N$. To take this into account, one adds to the energy Hamiltonian the term
\be
\label{27}
- \mu \hat N \; = \; - \mu \int \psi^\dgr(\br) \; \psi(\br) \; d\br 
\qquad ( T > T_c) \;   .
\ee
 
Below the condensation temperature $T_c$, there are two types of field operators $\eta$
and $\psi_1$. The condensate number of particles $N_0 = |\eta|^2$, by the Bogolubov-Ginibre
theorem is defined as a minimizer of a thermodynamic potential, say free energy. And the
average total number of particles $N$ remains also fixed. In that way, for a system with
broken gauge symmetry, there are two conditions fixing $N_0$ and $N$. As far as 
$N = N_0 + N_1$, there is no difference which pair of particle numbers to fix, either $N_0$
and $N$, or $N_1$ and $N$, or $N_0$ and $N_1$. In the latter case, the grand Hamiltonian is
defined by adding to the energy Hamiltonian the term
\be
\label{28}
- \mu_0 \hat N_0 -  \mu_1 \hat N_1 \; = \; - 
\mu_0 N_0 - \mu_1 \int \psi_1^\dgr(\br) \; \psi_1(\br) \; d\br 
\qquad ( T < T_c) \;    .
\ee
With two chemical potentials, no dilemma appears. The approach is conserving, 
self-consistent, respecting all thermodynamic relations, and it is gapless 
\cite{Yukalov_13,Yukalov_14,Yukalov_15,Yukalov_16,Yukalov_17}.  
    
\vskip 2mm
{\bf Condensate wave function}. When averaging the equation
\be
\label{29}
i \; \frac{\prt}{\prt t} \; \eta(\br,t) \; = \; 
\frac{\dlt H}{\dlt\eta^*(\br,t)}
\ee
over the vacuum of the field operator $\psi_1$, for which
\be
\label{30}
\psi_1(\br) \; | \; 0 \; \rgl_1 \; = \; 0 \; ,
\ee
it is straightforward to get the equation for the vacuum field
\be
\label{31}
i\; \frac{\prt}{\prt t} \; \eta(\br,t) \; = \; \left[ \; - \; 
\frac{\nabla^2}{2m} +  U - \mu_0 +  \int \Phi(\br-\br') \;
|\; \eta(\br',t) \; |^2 \; d\br' \; \right] \; \eta(\br,t) 
\ee
following from the equation
\be
\label{32}
 i\; \frac{\prt}{\prt t} \; \eta(\br,t) \; = \; _1\lgl \; 0 \; |
\; \frac{\dlt H}{\dlt\eta^*(\br,t)} \; | \; 0 \; \rgl_1 \;  .
\ee

The equation (\ref{31}) was advanced by Bogolubov \cite{Bogolubov_18} and since then 
republished numerous times (see, e.g., \cite{Bogolubov_3,Bogolubov_4,Bogolubov_5}). It 
was studied in many works, starting from \cite{Bogolubov_18,Gross_19,Gross_20,Gross_21,
Gross_22,Gross_23,Pitaevskii_24,Wu_25}.  

It is a very widespread fallacy to call this equation a ``mean-field approximation". This
equation is the vacuum-field equation. The general equation for the condensate wave 
function comes from the statistical averaging
\be
\label{33}
 i\; \frac{\prt}{\prt t} \; \eta(\br,t) \; = \; \left\lgl \;
\frac{\dlt H}{\dlt\eta^*(\br,t)} \; \right\rgl \; ,
\ee
which, as is easy to check, is rather different from (\ref{32}). Making in the equation
(\ref{33}) the Hartree-Fock-Bogolubov approximation gives the genuine mean-field equation,
also differing from (\ref{32}) \cite{Yukalov_12}. 

\vskip 2mm
{\bf Importance of symmetry-broken averages}. The other very common error is the omission 
of the symmetry-broken averages, like
\be
\label{34}
\sgm_k \; = \; \lgl \; a_k \; a_{-k}\; \rgl \;   .
\ee
One names this mistake ``Popov approximation". First of all, it is necessary to stress 
that Popov has never suggested such an unjustified trick, which is easy to confirm by 
attentively reading his works cited in that respect \cite{Popov_26,Popov_27,Popov_28}.    
Moreover, the omission of the anomalous averages, like $\sigma_k$ leads to the appearance
of unphysical singularities, incorrect first-order BEC transition, and a number of other 
inconsistences. This is because the anomalous averages appear simultaneously with the 
gauge symmetry breaking and BEC. Actually, they are the additional order parameters,
equally with the condensate wave function. This is why it is principally wrong omitting 
these averages \cite{Yukalov_12,Yukalov_29,Yukalov_30}.

\vskip 2mm
{\bf No thermodynamically anomalous fluctuations}. Dealing with the models describing 
realistic statistical systems in equilibrium, there should not appear fluctuations
making the system unstable. Otherwise, there can be no equilibrium or the model is not
realistic. Nevertheless, in many calculations there arise thermodynamically anomalous 
fluctuations. Such thermodynamically anomalous particle fluctuations are exhibited in
ideal Bose gases below the critical dimensionality. For example, the three-dimensional
uniform ideal Bose gas is unstable, displaying the compressibility divergence
\be
\label{35}
 \varkappa_T \; = \; \frac{m^2 T}{2\pi^3\rho^2} \; V^{1/3}\;  .
\ee
In the present case, this just means that the model is not realistic, which is not 
surprising, since in the real life there is no absolutely ideal gas. Always there are 
some, maybe very weak, interactions, e.g. due to collisions. And the existence of even 
infinitesimal interactions stabilizes the system. So, this is quite normal and realistic
situation. 

There is another cause for the appearance of divergences in the process of approximate 
calculations. Thus, when calculating particle fluctuations for a Bose-condensed system,
one employs the Bogolubov approximation, or hydrodynamic approximation, or 
Hartree-Fock-Bogolubov approximation. In all the cases, there is the same defect having
no physical reason but only the technical inconsistency that results in the occurrence of
exactly the same divergency as for the ideal gas.

The problem is that the ideal Bose gas is classified as a Gaussian-class model for which
the divergence in particle fluctuations is intrinsic. However the realistic Bose-condensed 
system pertains to the $XY$-class or $O(2)$-class. When one makes an approximation, such 
as Bogolubov, hydrodynamic, or Hartree-Fock-Bogolubov approximations, one reduces the 
model to quadratic form typical of the Gaussian-class models. Simultaneously with the 
simplification, one inherits the weak point of the Gausian model of producing divergent 
terms in the particle fluctuations. So, the appearance of these divergences, induced by
the model distortion, is not physical, but just technical. The cure of this is rather
simple, the divergent terms, induced by the model distortion, have to be subtracted. The
remaining parts give the thermodynamically normal compressibility
\be
\label{36}
\varkappa_T \; = \; \frac{1}{\rho m c^2 } \;   ,
\ee            
where $c$ is the sound velocity \cite{Yukalov_31}. Note that the same problem occurs 
when calculating the susceptibility of magnetic systems with continuous symmetry.   

\vskip 2mm
{\bf Conclusion}. This report discusses some of the difficult points arising in the 
theory of Bose-Einstein condensation. In the current literature, there are numerous
confusions related to these points. Here, we mention the ways helping to understand 
and to avoid these difficulties. The explanations are rather brief, following a report
style. A more detailed exposition will be published in a separate review-type paper.      

\vskip 2mm

{\bf Acknowledgment}. The author is grateful for useful discussions to V.S. Bagnato
and E.P. Yukalova

\end{document}